\documentclass{llncs}
\usepackage{times}
\usepackage{url}
\usepackage{latexsym}
\usepackage{array,multirow,graphicx}
\usepackage{epsfig}
\usepackage{sidecap}
\usepackage{amssymb}
\usepackage{enumitem}
\usepackage[misc]{ifsym}

\usepackage[top = 3.50cm, bottom = 3.50cm, left = 3.75cm, right = 4.00cm]{geometry}

\begin{document}

\title{A Large-Scale CNN Ensemble for Medication Safety Analysis}
\titlerunning{ADR detection}
\author{Liliya Akhtyamova\inst{1}, Andrey Ignatov\inst{2}\textsuperscript{\,(\Letter)}, John Cardiff\inst{1}}
\authorrunning{Liliya Akhtyamova et al.}
\tocauthor{Liliya Akhtyamova, Andrey Ignatov, John Cardiff}
\institute{Institute of Technology Tallaght, Ireland\\
\and ETH Zurich, Switzerland \\ \vspace{2mm}
\email{liliya.akhtyamova@postgrad.ittdublin.ie, andrey.ignatoff@gmail.com, john.cardiff@it-tallaght.ie}}

\maketitle
\begin{abstract}
Revealing Adverse Drug Reactions (ADR) is an essential part of post-marketing drug surveillance, and data from health-related forums and medical communities can be of a great significance for estimating such effects. In this paper, we propose an end-to-end CNN-based method for predicting drug safety on user comments from healthcare discussion forums. We present an architecture that is based on a vast ensemble of CNNs with varied structural parameters, where the prediction is determined by the majority vote. To evaluate the performance of the proposed solution, we present a large-scale dataset collected from a medical website that consists of over 50 thousand reviews for more than 4000 drugs. The results demonstrate that our model significantly outperforms conventional approaches and predicts medicine safety with an accuracy of 87.17\% for binary and 62.88\% for multi-classification tasks.

\end{abstract}
\section{Introduction}
Monitoring Adverse Drug Reactions (ADR)~--- unintended responses to a drug when it is used at recommended dosage levels, has a direct relationship with the public health and healthcare costs around the world. Side effects of medicines lead to 300 thousand deaths per year in the USA and Europe~\cite{Businaro2013WhyPharmacovigilance}, therefore revealing adverse drug reactions is of paramount importance for the government authorities, drug manufacturers, and patients. Data from the European Medicines Agency (EMA) shows that patients are not reporting side effects adequately through official channels, making it necessary to explore some different ways of ADR monitoring.

In this case, social media provides a substantial source of information that gives unique opportunities and challenges for detecting ADR using NLP techniques. It was shown that a large population of patients are actively involved in sharing and posting health-related information in various healthcare social networks \cite{Chou2009SocialCommunication.}, and thus the latter promise to be a powerful tool for monitoring ADR. However, the considered task still remains extremely challenging due to varying posts formats and the complexity of human language. Currently, deep neural networks have achieved impressive results on many NLP-related problems, and a particular success here have Convolutional Neural Networks. While having a large number of parameters, they demand massive datasets for efficient training, and the lack of large annotated text corpora made their application to ADR extraction task very limited. In this work, we eliminate this problem and propose an end-to-end solution for predicting drugs safety using an ensemble of CNNs.

The contributions of this paper are as follows:
(i) We present a large-scale ADR extraction dataset which we make available along with this paper.
(ii) We propose a CNN ensemble for tackling the problem of ADR binary and multi-classification that requires a minimal number of preprocessing steps and no hand-crafted features.
(iii) Our experimental results reveal that the proposed solution significantly outperforms baseline approaches and boosts the performance of the conventional CNN-based method.


\section{Related Research}

The earliest work on ADR extraction was \cite{LeamanTowardsNetworks}, where the authors investigated the potential of user comments for early detection of unknown ADR. This and the subsequent works were mainly focused on a limited number of drugs and were based on hand-designed features \cite{Sarker2015UtilizingAccess}. To tackle the problem of lack of investigation in this area some challenges were organized. One competition was Diegolab-2015, where the goal was to develop an algorithm for solving the problems of ADR classification and extraction. The teams showed competitive results on a difficult Twitter dataset, and the best performance was achieved by \cite{Sarker2016SocialTwitter} with 59\% F1-score for ADR class.

Another work contributing to the topic of ADR detection is~\cite{Gurulingappa2012ExtractionReports.}, where the authors used a publicly available ADE corpus for binary medical case reports classification and achieved an F-score of 77\% for documents with ADR. In~\cite{Sarker2015UtilizingAccess} the authors tried to combine ADE corpus with data from social media (DailyStrength, Twitter). In a number of papers ADR detection was considered from the position of sentiment analysis. In~\cite{Sharif2014DetectingFramework} the authors extracted semantic, sentiment, and affect features from two datasets (AskaPatient and Pharma tweets) and classified them using SVM with linear kernel~\cite{Joachims2006TrainingTimeb}. They split the datasets into two classes, defining ADR classification problem as a binary sentiment analysis problem, and achieved an accuracy of 78.20\% and 79.73\% for AskaPatient and Pharma datasets, respectively.

Though CNNs became a standard approach in many NLP-related tasks, only a few researches considered building a committee of several networks. In~\cite{Jan2016} the authors combined two separate CNNs with different architectures for the task of Twitter Sentiment Analysis. In~\cite{Sutskever2014} an ensemble of five CNNs was used for English to French machine translation, where it demonstrated superior results compared to a single network. In this work, we will show that larger committees can yield even better performance when a diverse set of CNNs is used.

\section{Method}


\subsection{Input Processing}

In our task, the input to the classification model has the form of a user text post $\mathbf{T}$ that is treated as an ordered sequence of words $\mathbf{T} = \{w_1,w_2,...,w_N\}$. First, plain words are mapped to their vector representations using a pre-trained word embedding model, which in our case is word2vec. The resulting representations are stacked together to form a single sentence matrix $\mathbf{M_T}$.
If the original text $\mathbf{T}$ consists of $N$ words and the dimensionality of word embeddings is $d$, this results in a $d\times N$ real-valued matrix which $i$-th column is a vector representation of the $i$-th word of the sentence. This matrix is then passed to a CNN and further steps are described below.

\subsection{CNN architecture}

\begin{figure}[t!]
  \centering
  \includegraphics[width=1\columnwidth]{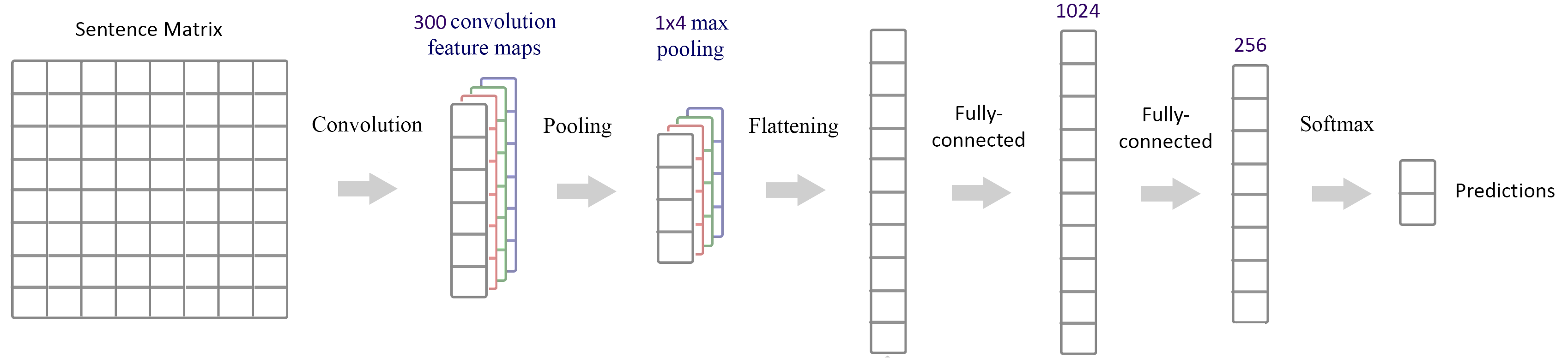}
  \caption{The overall architecture of the proposed CNN-based model}
  \label{fig:architecure}
\end{figure}

The architecture of our baseline CNN is presented in figure~\ref{fig:architecure}. It consists of one convolutional, one pooling and two fully-connected layers. The convolutional layer contains 300 filters of size $5\times d$, where $d$ is the dimensionality of word embeddings or the height of the sentence matrix. The number of neurons in the fully-connected layers is 1024 and 256. We use a dropout technique in these layers with dropout rate 0.2 to avoid overfitting. The CNN is trained to minimize cross-entropy loss function which is augmented with $l_2$-norm regularization of CNN weights, the parameters of the network are optimized with Adam algorithm.

\subsection{Ensemble of CNNs}

In this work we propose using a committee of up to 40 CNNs with various structural parameters that are trained on the same dataset. To guarantee the diversity of the models, we consider CNNs that have different number of convolutional filters (200 - 400), size of the convolutional filter (4-8) and dimension of word embeddings (200-300). The prediction of the committee is determined by a majority vote, and from the statistical viewpoint this combination of models is more powerful than a single one if sub-models are uncorrelated. If they are unbiased estimators of the true distribution, the combination will be still unbiased but with a reduced variance.

\section{Dataset Construction and Preprocessing}

In this paper, we present a public\footnote{please email the authors to get access to the dataset} dataset gathered from the popular health forum \textit{AskaPatient}\footnote{askapatient.com}, where people share their treatment experience. Each record in this forum is left by patients and consists of the following fields: drug name, drug rating, reason for taking this medication, side effects of the medication, comments, sex, age, duration and dosage, date added. User ratings are ranged from 1 to 5. In total \textit{AskaPatient} database contains 59912 publicly available reviews for about 4K drugs left during the last 5 years.

\smallskip

We consider the following two problems:

\vspace{-2mm}

\begin{itemize}[label=$\bullet$]
\item \textit{Binary ADR classification:} reviews with ratings 1-2 are labeled as negative (since they correspond to negative side-effects and contain ADR mention) and reviews with ratings 4-5 are labeled as positive (can be presumed as a positive medication experience). Posts with rating 3 are ignored as not truly positive, negative or neutral~\cite{Sharif2014DetectingFramework}.

\smallskip

\item \textit{Multi-class ADR classification:} we predict all five classes listed in user ratings
\end{itemize}

\vspace{-2mm}

Since different medical forums may not contain other fields except for explicit user comments, we use only this field in our experiments to make our system more general. We use 80\% of the dataset for training and 20\% for testing the model. The same proportion of train/test data is utilized for each class.

\section{Experiments}

\begin{table}[t!]
    \caption{Classification results for the binary classification problem}
\centering
    {\small
\begin{tabular}{|l|c|c|c|}
  \hline
   Method & ADR accuracy, \% & Non-ADR accuracy, \% & Overall accuracy, \% \\
  \hline
  Avg. embeddings + Logistic Reg. & 74.11 & 76.41 & 75.29 \\
  Bag-of-words + Logistic Reg. & 75.33 & 77.07 & 76.22 \\
  Avg. embeddings + Random Forest & 76.29 & 85.13 & 80.82 \\
  Bag-of-words + Random Forest & 77.94 & \textbf{85.19} & 81.65 \\
  Single CNN & 85.15 & 84.96 & 85.05 \\
  Ensemble of 20 CNNs & \textbf{89.3}8 & 84.92 & \textbf{87.17} \\
  \hline
\end{tabular}}
\label{tab:tabl1}
\end{table}

\subsection{Experimental Setup}

A word2vec neural language model is used to learn word embeddings on the \textit{AskaPatient} corpus. We consider a skipgram model with window size 5 and filter words with frequency less than 5. The dimensionality $d$ of word embeddings is set to 300 for a CNN-based model, while for an ensemble we additionally consider embeddings of size 200. Convolutional Neural Networks in both cases are trained for 20K iterations with a learning rate of 5e-4 and $l_2$-regularization set to 1e-2. 

\smallskip

To establish some baseline results on the presented dataset we have additionally implemented two algorithms commonly used in NLP. The first one is based on the \textit{bag-of-words} model that takes into account only the multiplicity of the appearing words, regarding neither the word order nor grammar. The text in this model is represented by a single vector with values indicating the number of occurrences of each vocabulary word in the text. To classify the
obtained vectors we use Logistic Regression and Random Forest classifiers. The second model is based on \textit{averaged  word embeddings}~-- instead of stacking the produced embeddings into a sentence matrix, their averaging is performed to obtain one vector of size $d$ that is further classified using the same algorithms. We use 500 trees for Random Forest and regularization term $C=0.01$ for Logistic Regression.

\subsection{Results and Discussion}

\begin{table}[b!]
    \caption{Classification results for the multi-classification problem}
\centering
    {\small
\begin{tabular}{|l|c|c|c|c|c|c|}
  \hline
   Method & Class 1 & Class 2 &  Class 3 &  Class 4 &  Class 5 & Overall, \% \\
  \hline
  Avg. embeddings + Logistic Reg. & 79.07 & 2.99 & 22.67 & 16.94 & 57.87 & 43.79 \\
  Bag-of-words + Logistic Reg. & 73.56 & 7.95 & 24.70 & 28.24 & 56.51 & 45.60 \\
  Bag-of-words + Random Forest & 72.26 & 13.22 & 28.17 & 31.63 & 67.37 & 49.59 \\
  Avg. embeddings + Random Forest & 81.49 & 26.44 & 33.93 & 35.94 & \textbf{74.09} & 57.85 \\
  Single CNN & 78.72 & 38.03 & 42.79 & \textbf{52.34} & 63.04 & 59.11 \\
  Ensemble of 40 CNNs & \textbf{83.86} & \textbf{38.08} & \textbf{49.25} & 50.85 & 69.18 & \textbf{62.88} \\
  \hline
\end{tabular}}
\label{tab:tabl2}
\end{table}

The summary results for our experiments are presented in tables~\ref{tab:tabl1} and \ref{tab:tabl2} for binary and multi-classification tasks respectively. These tables specify the predictive accuracy associated with each class of ADR for both baseline and proposed methods. As one can see, the basic CNN-based model demonstrates a notably stronger performance compared to the baseline approaches and reaches an accuracy of 85.05\% and 59.11\% on binary and multi-classification tasks accordingly. Using a proposed committee of 40 CNNs dramatically improves the results, outperforming a single CNN by over 2\% and 3.5\% of accuracy on both tasks.

\begin{figure}[t!]
  \centering
  \includegraphics[width=0.495\columnwidth]{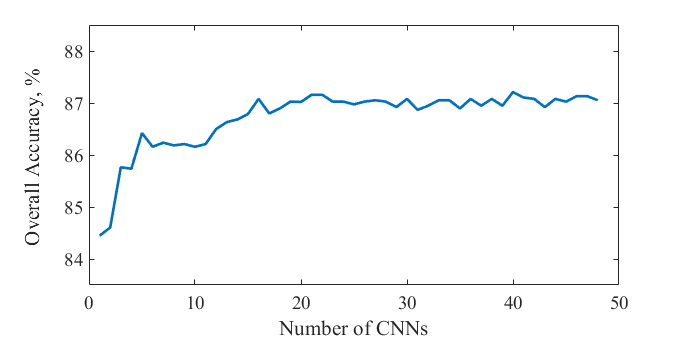}
  \includegraphics[width=0.495\columnwidth]{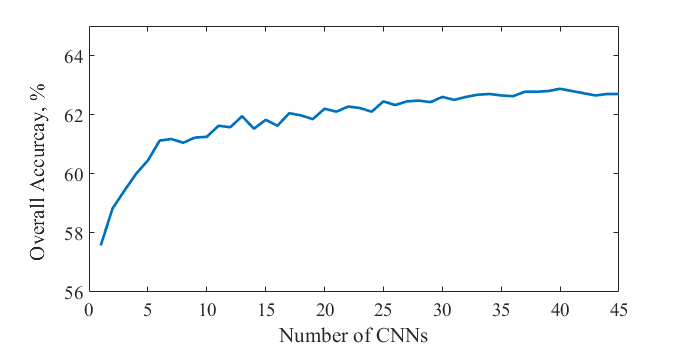}
  \caption{Dependency between the number of CNNs in ensemble and its total accuracy for the binary [left] and multi-classification [right] problems}
  \label{fg:ensemble}
\end{figure}

Figure~\ref{fg:ensemble} shows the dependency between the number of CNNs in the ensemble and its accuracy. First, the overall performance of ensemble grows as the number of CNNs increases, but then stabilizes at the threshold that is roughly equal to 20 CNNs for binary and 40 CNNs for multi-classification tasks; after this points we observe only slight fluctuations of the results.

The excellent performance of this model presumes that the proposed framework can facilitate enhanced detection of adverse drug events, with both better event recall and timelier identification. While tested in the context of adverse drug events, the framework is general to be applied to datasets from other domains, particularly when working with social media where the data volume is colossal and numerous sources of information exist. In such tasks it may significantly reduce annotation time and expenses.

\section{Conclusion and Future Work}

In this paper, we presented a large-scale ADR corpus crawled from a medical health forum. The corpus includes comments on drugs, user ratings and a number of other categories than can be used for a predictive model construction. We proposed an end-to-end solution that is based on a large ensemble of Convolutional Neural Networks, that in contrast to many previous works does not require any handcrafted features and data preprocessing. Our experimental findings show that the proposed model significantly outperforms baseline methods and introduces a large improvement to the standard CNN-based method.

We see several ways for the improvement of the existing solution. First of all, a wider set of models can be included in ensemble, particularly it can be reasonable to add Recurrent Neural Networks to the consideration. Secondly, a more sophisticated way of building a committee can be used, for instance bagging or boosting of the CNNs. Finally, we are planning to augment the existing dataset with data from other pharmaceutical forums and websites to create a reacher set of ADR mentions.

\bibliographystyle{splncs}

\end{document}